\documentclass[twocolumn,aps,showpacs,preprintnumbers]{revtex4}

\usepackage[latin1]{inputenc} 
\usepackage{graphicx}
\usepackage[intlimits]{amsmath}
\usepackage{amsxtra}
\usepackage{amssymb}
\usepackage{psfrag}
\usepackage{amstext}
\usepackage{dsfont}
\usepackage{amsfonts}

\newcommand{\erwart}[1]{\big\langle{#1}\big\rangle}
\newcommand{\bra}[1]{\left\langle{#1}\right|}
\newcommand{\ket}[1]{\left|{#1}\right\rangle}
\newcommand{\braket}[2]{\left\langle{#1}\right.\left|{#2}\right\rangle}

\newcommand{\mi}[1]{\min\left\{{#1}\right\}}
\newcommand{\e}{\operatorname{e}}

\newcommand{\fig}[1]{Fig.~\ref{#1}}
\newcommand{\eq}[1]{Eq.~(\ref{#1})}

\newcommand{\ketind}[2]{\sideset{}{_{#2}}{\mathop{\ket{#1}}}}

\begin{document}


\title{Quantum estimation of a damping constant}

\author{Hannah Venzl}\altaffiliation[Present address: ]{Max-Planck-Institut f\"ur Physik komplexer Systeme, N\"othnitzer Stra\ss e 38, 01187 Dresden, Germany}
\author{Matthias Freyberger}
\affiliation{Institut f\"ur Quantenphysik, Universit\"at Ulm, 89069 Ulm, Germany}


\begin{abstract}

We discuss an interferometric approach to the estimation of quantum mechanical damping. We study specific classes of entangled and separable probe states consisting of superpositions of coherent states. Based on the assumption of limited quantum resources we show that entanglement improves the estimation of an unknown damping constant.
\end{abstract}


\maketitle


\section{Introduction}

Nonclassical quantum states can be used in various ways to tune up measurement resolutions beyond the standard limits \cite{gerry,caves80}. 
The application of squeezed states of light in high-precision interferometry often serves as a prominent example \cite{caves81,bondurant84,xiao87,grangier87}. Here the improved precision results from quantum noise reduction due to a specific superposition of photon number states. Hence appropriately prepared quantum superpositions offer one way to beat standard quantum limits. However, the quantum mechanical state space allows us to consider more sophisticated, nonclassical resources. In fact one can exploit the entanglement advantage of coupled quantum systems. The corresponding nonclassical correlations have been shown to enhance the measurement precision \cite{sergienko03}. Examples can be found in interferometry \cite{yurke86}, in quantum ellipsometry \cite{toussaint04}, in atomic state spectroscopy \cite{wineland92}, in the improvement of frequency standards \cite{bollinger96,huelga97}, and in optical lithography \cite{boto00}.

All these examples of quantum metrology are influenced by decoherence effects. In general any coupling to an uncontrollable environment will diminish nonclassical enhancements. Therefore, it seems that decoherence has only a negative side when it comes to measurements based on quantum systems. However, counterexamples are known. Specific schemes for the generation of nonclassical correlations require decoherence \cite{plenio99}. Moreover, the fragility of quantum coherence effects can be used efficiently to determine absorptions in optical sensor technology \cite{scheel03}.

In the present contribution we shall describe an interferometric estimation process for a damping constant. If we choose nonclassical input states for the interferometer, the estimation quality will be considerably higher than in the corresponding classical case. Moreover, we shall discuss how much of the effect solely relies on quantum superposition and what can be associated to quantum entanglement.

The paper is organized as follows. We first introduce the model and the considered interferometric estimation scheme. Based on this scheme we shall then motivate different classes of nonclassical and classical probe states and discuss their properties in some detail.
The different estimation results will be compared in order to determine the role of quantum superposition and entanglement for the quality of the described estimation scenarios.

\section{Model}

\begin{figure}
\begin{center}
\includegraphics[width=.4\textwidth]
{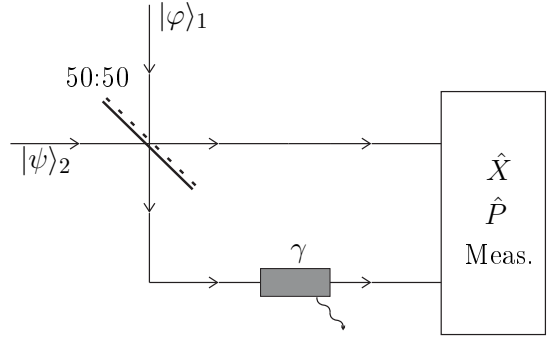}
\caption{The setup consists of a two-mode input state transformed by a 50:50 beam splitter. We assume unknown damping only in the first mode with the damping constant $\gamma$. A measurement apparatus (e.g., eight-port interferometer \cite{walker86,noh91}) registers eigenvalues of the EPR observables $\Hat{X}$ and $\Hat{P}$, \eq{EXP}.}
\label{Fsetup}
\end{center}
\end{figure}

The schematic setup is shown in \fig{Fsetup}.
It consists of two spatially separated modes $k=1,2$ of the electromagnetic field, described by the mode annihilation (creation) operators $\hat a_{k}\;(\hat a_{k}^\dagger)$. They are connected by a lossless 
50:50 beam splitter. The corresponding unitary transformation
\begin{equation}
\Hat{U}_\text{BS}=\e^{\frac{\pi}{4}(\hat a_1\hat a_2^\dagger-\hat a_1^\dagger\hat a_2)}
\label{Ebs}
\end{equation}
in general entangles a two-mode input state
\begin{equation}
\ket{\Psi}=\ket{\varphi}_1\otimes\ket{\psi}_2.
\label{Einput}
\end{equation}
The degree of entanglement will depend on their special form.
Behind this beam splitter, damping occurs in the first mode with the unknown damping constant $\gamma$.
Optical homodyne techniques can be used to measure quadrature observables
\begin{equation}
\hat x_{k}(\theta_{k})=\frac{1}{\sqrt{2}}(\hat a_k \e^{-i\theta_k}+\hat a^\dagger_k \e^{i\theta_k})
\label{Exp}
\end{equation}
for each mode $k$ with $\theta_{k} \;\in\; [0,2\pi)$. In particular, eight-port homodyning \cite{walker86, noh91} can be applied to determine the Einstein, Podolsky, and Rosen (EPR) observables \cite{epr}
\begin{equation}
\hat X=\frac{1}{\sqrt{2}}\big(\hat x_1-\hat x_2\big),\quad
\hat P=\frac{1}{\sqrt{2}}\big(\hat p_1+\hat p_2\big),
\label{EXP}
\end{equation}
where $\hat x_k\equiv\hat x_{k}(0)$ and $\hat p_k\equiv\hat x_{k}(\frac{\pi}{2})$.

Our aim is now to show that such an interferometric setup allows us to characterize the damping. In particular, we shall find that appropriate nonclassical input states, \eq{Einput}, which can be entangled by the beam splitter, will enhance the quality of this characterization \cite{scheel03}.
%

\subsection{Damping}

A common way to describe damping of an electromagnetic field mode is to consider its coupling to a reservoir of many modes, e.g., the radiation field of the vacuum. Here we will assume the usual linear coupling of the reservoir and the mode operators which models amplitude damping.

Behind the beam splitter this leads to the master equation \cite{louisell,carmichael}
\begin{equation}
\dot{\rho}(t)
=\sum_{k=1}^{2}\left(-i \omega [\Hat{a}^{\dagger}_{k}\Hat{a}_{k},\rho]\right)+\gamma{\cal{L}}_{1}\rho,
\label{Emaster}
\end{equation}
for a two-mode density operator $\rho$,
under Born-Markov approximation in the zero temperature limit, where $\omega$ is the frequency of both modes.
The dissipative evolution of mode 1 is described by the Lindblad term
\begin{equation}
\gamma{\cal{L}}_{1}\rho\equiv
\frac{\gamma}{2} \;\left( 2 \Hat{a}_{1} \rho \Hat{a}_{1}^{\dagger} -\Hat{a}_{1}^{\dagger} \Hat{a}_{1} \rho -\rho \Hat{a}_{1}^{\dagger} \Hat{a}_{1}\right),
\label{EmasterWW}
\end{equation}
where $\gamma$ is the damping constant.
Since we are considering optical modes, the zero temperature assumption is reasonable.

For an initial density matrix element $\ket{\alpha }_1\bra{\alpha' } \otimes\ket{\beta }_2\bra{\beta' }$  in a coherent state basis one finds \cite{walls85,phoenix90}
\begin{equation}
\begin{split}
&\big(\ket{\alpha }_1\bra{\alpha' }\otimes\ket{\beta }_2\bra{\beta' }\big) (\kappa)\\
&\equiv\e^{\kappa{{\cal{L}}}_1}\big(\ket{\alpha }_1\bra{\alpha' }\otimes\ket{\beta }_2\bra{\beta' }\big)\\
&=\braket{\alpha'}{\alpha}^{1-e^{-\kappa}}
\ket{\alpha e^{-\kappa/2}}_1\bra{\alpha' e^{-\kappa/2}}\otimes\ket{\beta }_2\bra{\beta'}
\end{split}
\label{Edensityterms}
\end{equation}
for the solution
of \eq{Emaster} in the interaction picture. Here we have introduced the scaled damping constant $\kappa\equiv\gamma t$.

\section{Estimation of an unknown damping constant}\label{Cest}

After this discussion of our model, we shall now introduce the estimation scheme.
We then derive an equation which allows us to compare the fidelity of the estimation using different input states, \eq{Einput}.
In particular we can choose input states which lead to nonclassical correlations between the modes of the interferometer.

As mentioned before, we assume damping only in mode 1, or in other words, the damping in mode 2 is supposed to be much weaker than the damping in mode 1.
The second mode is thus basically considered as a perfect reference mode.
However, the more general case of independent damping in both modes can be analyzed analogously and leads to very similar results.
To derive an equation for the average error obtained when estimating the unknown scaled damping constant $\kappa$, we assume that we perform $N$ measurements of an observable $\Hat{\Omega}$, where $\Hat{\Omega}$ is in general a function of the EPR observables $\Hat{X}$ and $\Hat{P}$, \eq{EXP}. The $N$ measurements are supposed to be independent and carried out using identically prepared probe states, \eq{Einput}. The outcome of a single measurement run therefore consists of $N$ eigenvalues $\{\Omega_1,...,\Omega_N\}\equiv\{\Omega_i\}$ of the observable $\Hat{\Omega}$.

In order to find a simple expression for the estimated parameter $\kappa$ and under the assumption of weak coupling and short damping times, we linearize the expectation value
\begin{equation}
\begin{split}
\erwart{\Hat{\Omega}}(\kappa)&=\text{Tr}\left\{\rho(\kappa\equiv \gamma t)\;\Hat{\Omega}\right\}\\
&\cong \erwart{\Hat{\Omega}}(0)+\frac{d\erwart{\Hat{\Omega}}(\kappa)}{d\kappa}\bigg|_{\kappa=0} \kappa,
\end{split}
\label{Eexpectvalue}
\end{equation}
which contains $\kappa$ via the time evolved density operator $\rho(\kappa)$.
We can then write $\kappa$ as a linear function of $\erwart{\Hat{\Omega}}(\kappa)$
\begin{equation}
\kappa\cong c_0+c_1 \erwart{\Hat{\Omega}}(\kappa),
\label{Ekappa}
\end{equation}
where the coefficients $c_m$ follow directly from \eq{Eexpectvalue}.
Based on the $N$ measurement results $\Omega_i$ of a single run we will estimate the scaled damping constant
\begin{equation}
\kappa_\text{est}(\{\Omega_i\})\equiv c_0+c_1 \frac{1}{N}\sum_{i=1}^{N}\Omega_i.
\label{Eest}
\end{equation}
As long as the linearization is valid this choice guarantees that $\kappa_\text{est}(\{\Omega_i\})\to\kappa$ for $N\to\infty$. We note, however, that this simplified approach can be refined. One possibility would be to obtain $\kappa_\text{est}(\{\Omega_i\})$ from the optimization of an appropriate likelihood function \cite{harney}. In order to bring out the nonclassical features of our estimation scheme it is sufficient to work with the simple linearization.

\subsection{Average estimation error}

Considering identical experiments, the estimated values $\kappa_\text{est}(\{\Omega_i\})$, \eq{Eest}, will still vary from run to run according to a distribution  
which depends on the input states and has the expectation value $\kappa$.
If we want to compare certain estimation scenarios based on different input states, we thus have to find a quantity defining the fidelity of a scenario. A simple possibility is the average width
\begin{equation}
\Delta\kappa^2\equiv\overline{\big(\kappa-\kappa_\text{est}(\{\Omega_i\})\big)^2},
\label{Edelta}
\end{equation}
as a function of the parameters defining our input states. The averaging is performed over infinitely many runs. Note that, for a single run, the quantity $\big(\kappa-\kappa_\text{est}(\{\Omega_i\})\big)^2$ depends on the special set $\{\Omega_1,...,\Omega_N\}$ of measured values; see \eq{Eest}.

As mentioned above, the individual measurements in each run are independent and carried out using identically prepared states. We can therefore  write
\begin{equation}
\Delta\kappa^2
=\left(\prod_{i=1}^{N}\int\limits_{-\infty}^{\infty}\!d\Omega_i  \;W(\Omega_i,\kappa)\right) \;\big(\kappa-\kappa_\text{est}(\{\Omega_i\})\big)^2
\end{equation}
with the probability distribution
\begin{equation}
W(\Omega, \kappa)=\bra{\Omega}\rho(\kappa)\ket{\Omega}
\end{equation}
given by the eigenstates $\ket{\Omega}$ of the observable $\Hat{\Omega}$ and the two-mode density operator $\rho(\kappa)$. We note
\begin{equation}
\overline{\Omega^n}=\int\limits_{-\infty}^{\infty}\!d\Omega \;W(\Omega,\kappa)\;\Omega^n=\erwart{\Hat{\Omega}^n}(\kappa),
\label{Eqmev}
\end{equation}
which is consistent with $\overline{\kappa_\text{est}(\{\Omega_i\})}=\kappa$.

The average estimation error, \eq{Edelta}, then reads
\begin{equation}
\begin{split}
\Delta\kappa^2&=\frac{c_1^2}{N}\;\left( \erwart{\hat \Omega^2}(\kappa)-\erwart{\hat \Omega}^2(\kappa)\right)\\
&=\frac{1}{N}\;\Delta \Omega^2(\kappa)\; \left(\frac{d\erwart{\Hat{\Omega}}(\kappa)}{d\kappa}\bigg|_{\kappa=0}\right)^{-2},
\end{split}
\label{GDelta}
\end{equation}
where we have introduced the variance
\begin{equation}
\Delta \Omega^2(\kappa)=\erwart{\hat \Omega^2}(\kappa)-\erwart{\hat \Omega}^2(\kappa)
\label{Evariance}
\end{equation}
and used Eqs. (\ref{Eexpectvalue}) and (\ref{Ekappa}).

Considering, for example, an estimation based on the observable $\Hat{\Omega}=\Hat{X}$,  suitable probe states should have minimal noise in the two-mode $\Hat{X}$ quadrature and the initial steepness $[d\langle\Hat{X}\rangle(\kappa)/d\kappa]|_{\kappa=0}$ of the first moment should be large.

\subsection{Limited resources}

Since we want to compare the behavior of different estimation scenarios we have to define which scenarios are considered comparable. In order to do so, we first assume that the special form of the input states $\ket{\Psi}$, \eq{Einput}, has no influence on the preparation complexity. Furthermore, the beam splitter is a passive element and hence does not involve any further resources to entangle input states.
We therefore consider two scenarios comparable if the average total photon number in one run,
\begin{equation}
n_\text{tot}\equiv N \,\erwart{n},
\label{Entot}
\end{equation}
with the average photon number in one measurement,
\begin{equation}
\erwart{n}\equiv \bra{\Psi}(\hat n_1+\hat n_2)\ket{\Psi} =\erwart{\hat a_1^\dagger \hat a_1}+\erwart{\hat a_2^\dagger \hat a_2},
\label{En}
\end{equation}
is the same. We do not, however, fix how many measurements are performed in one run or, in other words, how many photons are contained in one measurement of $\hat \Omega$. Using this idea of limited resources, the average quadratic error, \eq{GDelta},
\begin{equation}
\Delta\kappa^2=\frac{\erwart{n}}{n_\text{tot}}\;\Delta \Omega^2(\kappa)\; \left(\frac{d\erwart{\Hat{\Omega}}(\kappa)}{d\kappa}\bigg|_{\kappa=0}\right)^{-2},
\label{Eerror}
\end{equation}
has to be minimized as a function of 
the parameters describing the probe states for a given number $n_\text{tot}$.


\section{Probe states}

Our aim is to compare classical and nonclassical estimation scenarios based on the error given by \eq{Eerror}. We are therefore interested in finding appropriate observables $\hat \Omega$, which are in general functions of $\hat{X}$ and $\hat{P}$, \eq{EXP}, and suitable input states, \eq{Einput}. Guided by the expression for the average error we have to select those input states which show minimal noise in the corresponding observable.

In order to work out the role of entanglement for the estimation of an unknown damping constant we regard both entangled and separable superposition states and compare them to what we will refer to as classical probe states.  We will therefore discuss our setup, see \fig{Fsetup}, with and without the beam splitter respectively and define suitable probe states for each case.

\subsection{Entangled probe states}\label{ent}

To motivate the first set of probe states, we first consider our setup, see \fig{Fsetup}, with the beam splitter in place but without any damping. So, in general, we will entangle the input states.
For the identification of the parameters controlling the noise $\Delta \Omega^2$, \eq{Evariance}, we have to examine the probability distribution
\begin{equation}
W(X,P)\equiv\big|\bra{X,P}\Hat{U}_\text{BS}\ket{\Psi}\big|^2,
\label{Eprob}
\end{equation}
which is determined by the input states transformed by the beam splitter and by the simultaneous eigenstates $\ket{X,P}$ of the commuting observables $\hat{X}$ and $\hat{P}$, \eq{EXP}. We note that those eigenstates can be written as
\begin{equation}
\ket{X,P}=\Hat{U}_\text{BS}\ket{p=P}_1\otimes \ket{x=-X}_2
\end{equation}
using the quadrature eigenstates $\ketind{p}{1}$ and $\ketind{x}{2}$ of $\hat p_1$ and $\hat x_2$. For the separable input state $\ket{\Psi}$, \eq{Einput}, we then find
\begin{equation}
W(X,P)= \big|\braket{p=P}{\varphi}_1 \big|^2\;\big|\braket{x=-X}{\psi}_2\big|^2
\label{EprobWITH}
\end{equation}
for the probability distribution of the observables $\hat{X}$ and $\hat{P}$, \eq{EXP}. Note that each input mode only affects the distribution of one of the considered observables. The noise can be controlled independently.

As shown before, for the estimation we have to find states that allow us to minimize the noise in the measured observable $\hat{\Omega}$.
From \eq{EprobWITH} we can see that ideal input states should consist of two quadrature states.
Since quadrature states have infinite energy, i.e., average photon number,  we rely on a more physical class of input states which can approximate those states to some extent. Clearly the first choice would be appropriately squeezed states. However, it has been shown, see, e.g. \cite{schleich91}, that by simply superposing coherent states, noise can be suppressed in single-mode quadratures $\hat x_k(\theta_k)$, \eq{Exp}. When we finally add damping, this has the advantage that a simple analytic solution to the two-mode master equation, \eq{Emaster}, is known; see \eq{Edensityterms}.

\begin{figure}
\begin{center}
\includegraphics[width=.4\textwidth]
{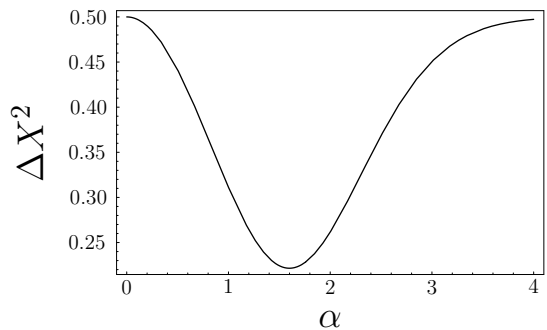}
\caption{Variance $\Delta X^2$, \eq{EvarianceMITX}, as a function of the parameter $\alpha$. The fluctuations are determined by the input mode 2 only.}
\label{FvarianceMIT}
\end{center}
\end{figure}

Moreover, we restrict our analysis to the case of measuring the observable $\hat\Omega=\hat X$, \eq{EXP}, only, since the principle can already be obtained from this reduced problem.
We thus conveniently consider superpositions of coherent states in mode 2 and vacuum noise in mode 1. The first class of input states will therefore be of the form
\begin{equation}
\ket{\Psi_\text{I}}=\ket{0}_1\otimes{\cal{N}}_{\alpha}\;
\left(\ket{i\frac{\alpha}{2}}_2+\ket{-i\frac{\alpha}{2}}_2\right),
\label{EinputWITH}
\end{equation}
properly normalized with the constant $\cal{N}_\alpha$ and $\alpha \in \mathbb{R} $. For those states, the probability distribution, \eq{EprobWITH}, has the simple form
\begin{equation}
W(X,P)
=\frac{2\,\left|{\cal N}_\alpha\right|^2}{\pi} \e^{-X^2-P^2}\left(1+\cos{[\sqrt{2}\,\alpha X]}\right).
\label{Eprobwith}
\end{equation}
It can then be shown in a straightforward way that the variance $\Delta X^2$ is given by \cite{schleich91}
\begin{equation}
\Delta X^2=\frac{1}{2}-\frac{\alpha^2}{2\;(1+e^{\alpha^2/2})}.
\label{EvarianceMITX}
\end{equation}
See \fig{FvarianceMIT}.
The minimum, $\mi{\Delta X^2}=0.22$, is below the classical value of $0.5$ and is reached for $|\alpha|=1.6$. The corresponding variance in $P$ stays constant, $\Delta P^2=0.5$.

For the states given by \eq{EinputWITH} we further find that the distribution $W(X,P)$, \eq{Eprobwith}, is localized at the origin of $X$-$P$ phase space; that is, $\erwart{\Hat{X}}=\erwart{\Hat{P}}=0$. Since we want to see a strong influence of the damping on first moments, see \eq{Eerror}, we have to displace this distribution in $X$-$P$ phase space. We note the relation
\begin{equation}
\bra{X,P}\hat D_1(X_0+iP_0)=\e^{i\Upsilon}\bra{X-X_0,P-P_0},
\end{equation}
where the phase $\Upsilon$ is a function of $X,P,X_0$, and $P_0$ and
\begin{equation}
\hat D_k(\delta)\equiv \e^{\delta \,\hat a^\dagger_k- \delta^* \hat a_k}
\end{equation}
is the single-mode displacement operator of mode $k$.
Moreover, the beam splitter transformation yields
\begin{equation}
\begin{split}
 & \;\Hat{D}_1(X_0+iP_0)\;\Hat{U}_\text{BS} \\
&=\Hat{U}_\text{BS}\Hat{D}_1\Big(\frac{X_0+iP_0}{\sqrt{2}}\Big)\;\Hat{D}_2\Big(-\frac{X_0+iP_0}{\sqrt{2}}\Big).
\end{split}
\end{equation}
Applying this combined displacement transformation to an arbitrary input state, \eq{Einput}, will therefore not change the form of the corresponding probability distribution $W(X,P)$, \eq{EprobWITH}, but simply shift it to the two-mode phase-space point
\begin{equation}
\erwart{\Hat{X}}=X_0,\quad\erwart{\Hat{P}}=P_0.
\end{equation}

Since we are only considering a measurement of the observable $\hat X$, \eq{EXP}, we displace the states according to
\begin{equation}
\ket{\Phi_\text{I}}\equiv \Hat{D}_1\Big(\frac{X_0}{\sqrt{2}}\Big)\;\Hat{D}_2\Big(-\frac{X_0}{\sqrt{2}}\Big)\ket{\Psi_\text{I}}.
\label{GinputWith}
\end{equation}
In other words, we increase the intensity in both input modes by the same amount.
These input states are then transformed by the beam splitter.
We will therefore use the states $\ket{\Phi_\text{I}}$ entangled by the beam splitter as suitable probe states for the estimation of the unknown damping.

\subsection{Separable probe states}\label{sep}

In order to distinguish the role of entanglement from the role of quantum superposition for the estimation of unknown damping, we now consider our setup, see \fig{Fsetup}, without the beam splitter. Similar to the procedure in the preceding paragraph, we will define suitable separable probe states, guided by the interferometric principle, \fig{Fsetup}, without damping.
The probability distribution 
\begin{equation}
W(X,P)\equiv\big|\braket{X,P}{\Psi}\big|^2
\label{Eprobwithout}
\end{equation}
is in this case directly given by the input state $\ket{\Psi}$, \eq{Einput}, and will not factorize like \eq{EprobWITH}.
In principle, the ideal minimal noise states are now given by eigenstates of the observables $\hat{X}$ and $\hat{P}$, \eq{EXP}, as can be seen from \eq{Eprobwithout}.
Since those states cannot be prepared experimentally and above all are not separable, we again try to approximate them with superpositions of coherent states.
The restriction to a measurement of the observable $\hat\Omega=\hat X$, \eq{EXP}, leads us to the symmetric input states
\begin{equation}
\begin{split}
\ket{\Psi_\text{II}}={\cal{N}}_\alpha^2\;&\left(\ket{i\frac{\alpha}{2}}_1+ \ket{-i\frac{\alpha}{2}}_1\right)
\otimes
\left(\ket{i\frac{\alpha}{2}}_2+ \ket{-i\frac{\alpha}{2}}_2\right)
\end{split}
\label{EinputWITHOUT}
\end{equation}
with the normalization ${\cal{N}}_\alpha^2$ and $\erwart{\Hat{X}}=\erwart{\Hat{P}}=0$. The form of the input states $\ket{\Psi_\text{II}}$ can be understood by noting that $\hat X$, \eq{EXP}, consists of a sum of one-mode quadratures $\hat x_{k}$. Therefore, noise has to be suppressed now in both input quadratures.

\begin{figure}
\begin{center}
\includegraphics[width=.4\textwidth]
{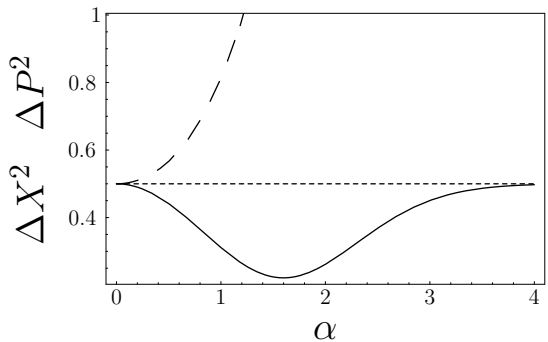}
\caption{Variances $\Delta P^2$, \eq{EvarianceOHNEP} (dashed line) and $\Delta X^2$, \eq{EvarianceOHNEX} (solid line) as a function of the parameter $\alpha$. The dotted line represents vacuum noise. It can easily be seen that reduced noise in $X$ strongly affects the width in $P$.}
\label{FvarianceOHNE}
\end{center}
\end{figure}

The variances are given by
\begin{equation}
\Delta X^2=\frac{1}{2}-\frac{\alpha^2}{2\;(1+e^{\alpha^2/2})}
\label{EvarianceOHNEX}
\end{equation}
and
\begin{equation}
\Delta P^2=\frac{1}{2}+\frac{\alpha^2\;e^{\alpha^2/2}}{2\;(1+e^{\alpha^2/2})},
\label{EvarianceOHNEP}
\end{equation}
respectively. States of the form $\ket{\Psi_\text{II}}$, \eq{EinputWITHOUT},  therefore allow for reduced noise in the observable $\hat{X}$, \eq{EXP}, to the same extent as the entangled probe states $\hat U_\text{BS}\ket{\Psi_\text{I}}$, \eq{EinputWITH}, discussed in the preceding section; see \eq{EvarianceMITX}.
However, as expected, reducing the noise in $X$ leads to a broadening of the probability distribution, \eq{Eprobwithout}, in $P$; see \fig{FvarianceOHNE}. The minimal value of $\Delta X^2=0.22$, for example, corresponds to $\Delta P^2=1.5$ which lies above the classical value of $0.5$.
The restriction to separable states leads to the feature that only one of the observables $\hat{X}$ or $\hat{P}$, \eq{EXP}, can be chosen to show reduced noise at a time \cite{duan00}.

In analogy to the discussion in the preceding paragraph we can define
\begin{equation}
\ket{\Phi_\text{II}} \equiv \Hat{D}_1(X_0) \ket{\Psi_\text{II}}
\label{GinputWithout}
\end{equation}
as the final displaced class of separable probe states with
$\erwart{\Hat{X}}=X_0$ and $\erwart{\Hat{P}}=0$.

The two classes of probe states $\ket{\Phi_\text{I}}$, \eq{GinputWith}, with beam splitter and $\ket{\Phi_\text{II}}$, \eq{GinputWithout}, without beam splitter, respectively, will make it possible to analyze separately the influence of quantum superposition and entanglement on the quality of an estimation scenario in our scheme.

Finally, we also compare these two classes of states to the most classical case with just one coherent state in each of the two input modes.
Those states can be obtained by just setting $\alpha=0$ in Eqs. (\ref{GinputWith}) and (\ref{GinputWithout}).

Depending on whether the interferometer is used with or without beam splitter this leads to
\begin{equation}
\ket{\Phi_\text{III}} =\ket{\frac{X_0}{\sqrt{2}}}_1\otimes\ket{\frac{-X_0}{\sqrt{2}}}_2
\label{GinputC1}
\end{equation}
and
\begin{equation}
\ket{\Phi_\text{IV}} = \ket{X_0}_1\otimes\ket{0}_2,
\label{GinputC2}
\end{equation}
respectively.
Before we come to the actual estimation, we shall now analyze the important features of our probe states in a bit more detail. In particular we shall outline how they are influenced by damping.

\subsection{Average photon numbers}\label{photon}

We first note the average photon numbers in one measurement, \eq{En}, for the different classes of input states. We obtain 
\begin{equation}
\begin{split}
\erwart{n_\text{I}}&\equiv \bra{\Phi_\text{I}}(\hat n_1+\hat n_2)\ket{\Phi_\text{I}}\\
&=X_0^2+\frac{1}{4} \alpha^2 \tanh{\frac{\alpha^2}{4}}
\end{split}
\end{equation}
for the input states $\ket{\Phi_\text{I}}$, \eq{GinputWith}, and 
\begin{equation}
\erwart{n_\text{II}}=X_0^2+\frac{1}{2} \alpha^2 \tanh{\frac{\alpha^2}{4}}
\end{equation}
for the probe states $\ket{\Phi_\text{II}}$, \eq{GinputWithout}, respectively. To generate the same amount of noise reduction more photons are needed in the separable than in the entangled case. The number of photons is identical for both classes of classical states, $\erwart{n_\text{III}}=\erwart{n_\text{IV}}=X_0^2$.

\subsection{Influence of damping}\label{time}

\begin{figure}
\begin{center}
\includegraphics[width=.4\textwidth]
{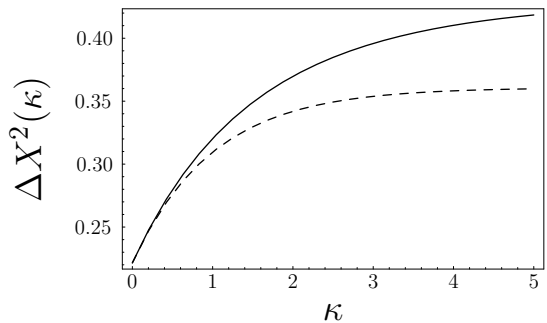}
\caption{Variance $\Delta X^2_\text{I}(\kappa)$, \eq{EvarianceWithKappa}, of the entangled probe states (solid line) and the variance $\Delta X^2_\text{II}(\kappa)$, \eq{EvarianceWithoutKappa}, of the separable superposition state (dashed line) as a function of the scaled damping constant $\kappa$. Here the parameter $\alpha$ is chosen such that the states initially show maximal noise reduction.}
\label{FdeltaXkappa}
\end{center}
\end{figure}

For estimating an unknown scaled damping constant, the effect of the damping on the probe states of course plays the crucial role for our considerations. According to \eq{GDelta}, two factors are of major importance: The variance $\Delta X^2(\kappa)$, \eq{Evariance}, and the first moment $\erwart{\hat X}(\kappa)$, \eq{Eexpectvalue}. For each of the considered probe states the first moment is given by
\begin{equation}
\erwart{\hat X}(\kappa)=X_0\, \e^{-\kappa/2},
\end{equation}
which basically shows the dissipative part of the damping mechanism.
For the evolution of the variance, in which the decoherence is expressed, we find
\begin{equation}
\Delta X^2_\text{I}(\kappa)=\frac{1}{2}-\frac{\alpha^2\, (1+\e^{-\kappa/2})^2}{8 \,(1+\e^{\alpha^2/2})}
\label{EvarianceWithKappa}
\end{equation}
and
\begin{equation}
\Delta X^2_\text{II}(\kappa)=\frac{1}{2}-\frac{\alpha^2\, (1+\e^{-\kappa})}{4\, (1+\e^{\alpha^2/2})},
\label{EvarianceWithoutKappa}
\end{equation}
respectively; see \fig{FdeltaXkappa}.
It is clearly visible that the entangled state $\hat U_\text{BS}\ket{\Phi_\text{I}}$, \eq{GinputWith}, loses the noise reduction faster than the separable superposition $\ket{\Phi_\text{II}}$, \eq{GinputWithout}. However, due to the fact that we are considering asymmetric damping, i.e., only damping in the first mode, also the final values of the variances are not equal. Clearly, the variances of the classical coherent states stay constant, i.e., $\Delta X^2_\text{III}(\kappa)=\Delta X^2_\text{IV}(\kappa)=0.5$.

\section{Estimation Results}\label{results}

We shall now compare the different estimation scenarios based on the basic assumptions described in Sec. \ref{Cest}. We are thus searching numerically for the minimum of $\Delta \kappa^2$, \eq{Eerror}, under the constraint of a constant photon number $n_\text{tot}$, \eq{Entot}, for the various discussed probe states. In order to compare the estimation results of nonclassical and classical input states, we define the relative improvement
\begin{equation}
\delta_\text{I/II}\equiv\frac{\Delta\kappa^2_\text{III}-\Delta\kappa^2_\text{I/II}}{\Delta\kappa^2_\text{III}}.
\label{Erelative}
\end{equation}
Here $\delta_\text{I}$ ($\delta_\text{II}$) denotes the relative improvement of states of the form $\hat U_\text{BS}\ket{\Phi_\text{I}}$, \eq{GinputWith}, ($\ket{\Phi_\text{II}}$, \eq{GinputWithout}) compared to classical input states $\ket{\Phi_\text{III}}$, \eq{GinputC1}. The comparison to classical input states $\ket{\Phi_\text{IV}}$, \eq{GinputC2}, is of course equivalent.

\begin{figure}
\begin{center}
\includegraphics[width=.4\textwidth]
{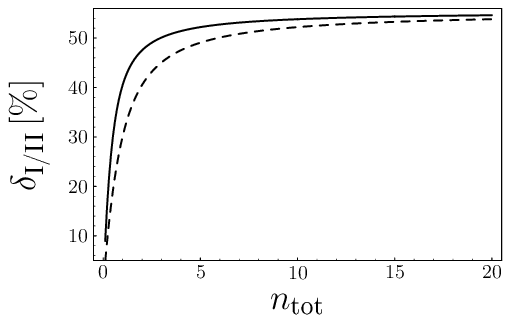}
\caption{Relative improvement $\delta_\text{I/II}$, \eq{Erelative}, in the estimation when comparing nonclassical probe states to classical probe states. The dashed line represents $\delta_\text{II}$, the comparison of separable superposition states $\ket{\Phi_\text{II}}$, \eq{GinputWithout}, to classical input states $\ket{\Phi_\text{III}}$, \eq{GinputC1}, and the solid line represents $\delta_\text{I}$, the comparison of entangled superpositions $\hat U_\text{BS}\ket{\Phi_\text{I}}$, \eq{GinputWith}, to classical input states.}
\label{FdeltaKappa}
\end{center}
\end{figure}

The obtained numerical results are shown in \fig{FdeltaKappa} for up to $n_\text{tot}=20$, using a scaled damping constant of $\kappa=0.01$.
It is clearly visible that, independent of the number of available photons, applying either separable or entangled quantum superpositions leads to a higher quality of the estimation. By using nonclassical probe states for the estimation, the error can be reduced by more than 50\% when compared to classical input states. In addition, for our class of entangled superposition states $\hat U_\text{BS}\ket{\Phi_\text{I}}$, \eq{GinputWith}, the error is smaller than for the separable superpositions $\ket{\Phi_\text{II}}$, \eq{GinputWithout}.

For just coherent states in each mode the error $\Delta \kappa^2$ scales like $n_\text{tot}^{-1}$. However, for both separable and entangled superpositions the scaling depends on the number of available photons.
Only from about $n_\text{tot}>5$ on the behavior is again governed by $\Delta \kappa^2\propto n_\text{tot}^{-1}$, which can be seen in the saturation of $\delta_\text{I}$ and $\delta_\text{II}$ in \fig{FdeltaKappa}.
This effect can be explained by noting that for a fixed photon number the average quadratic error $\Delta \kappa^2$, \eq{Eerror}, is influenced by both the variance $\Delta X^2(\kappa)$ and the initial steepness of the first moment $[d\langle\Hat{X}\rangle(\kappa)/d\kappa]\big|_0$.
For small photon numbers the suppression of noise in $\hat X$, \eq{EXP}, has not yet reached its maximal value. For larger photon numbers all the additional photons are then, like for classical states, being used to increase the intensity and thus the initial steepness of the first moment $[d\langle\Hat{X}\rangle(\kappa)/d\kappa]\big|_0$.
As discussed in Sec. \ref{photon}, less photons are needed to generate the same suppression of noise when using entangled instead of separable superposition states. This leads to the smaller average quadratic error for entangled states.

It is also interesting to note that as soon as the possibility of noise reduction is used it is on average best to use all available photons in one measurement. This can again be understood by recalling that a given fraction of photons is always needed for the suppression of noise.

We conclude by stating that, in our interferometric scheme, applying two-mode quantum superpositions improves the estimation of an unknown damping constant. Entangling those superpositions by a beam splitter even further improves this estimation. However, the main improvement when compared to an estimation with classical input states is given by the reduced noise in a quantum superposition state.

We further note that our estimation scheme can be considerably refined. Considering, for example, squeezed states or different observables could lead to an additional improvement of the estimation. By applying an adaptive estimation scheme \cite{denot06} one could possibly obtain a different scaling of the error $\Delta \kappa^2$, \eq{Eerror}, with the average total photon number $n_\text{tot}$, \eq{Entot}, even for larger photon numbers.


\begin{acknowledgments}

We thank Alexander Wolf for carefully reading the manuscript and we acknowledge the financial support by the graduate school ``Mathematical Analysis of Evolution, Information and Complexity'' at Ulm University.

\end{acknowledgments}



\end{document}